\documentclass[twocolumn,twoside,slac]{revtex4}
\usepackage{graphicx}
\usepackage{fancyhdr}
\def\dotsim{\stackrel{\cdot}{\sim}}

\begin{document}

\title{Likelihood Inference in the Presence of Nuisance Parameters}

%

\author{N. Reid, D.A.S. Fraser}
\affiliation{Department of Statistics, University of Toronto, Toronto Canada M5S 3G3}
%

\begin{abstract}
We describe some recent approaches to likelihood based inference in the presence of nuisance parameters.  Our approach is based on plotting the likelihood function and the $p$-value function, using recently developed third order approximations.  Orthogonal parameters and adjustments to profile likelihood are also discussed.  Connections to classical approaches of conditional and marginal inference are outlined.

\end{abstract}

\maketitle

\thispagestyle{fancy}


\section{INTRODUCTION}

We take the view that the most effective form of inference is provided by the observed likelihood function along with the associated $p$-value function.  In the case of a scalar parameter the likelihood function is simply proportional to the density function.  The $p$-value function can be obtained exactly if there is a one-dimensional statistic that measures the parameter.  If not, the $p$-value can be obtained to a high order of approximation using recently developed methods of likelihood asymptotics.  In the presence of nuisance parameters, the likelihood function for a (one-dimensional) parameter of interest is obtained via an adjustment to the profile likelihood function.  The $p$-value function is obtained from quantities computed from the likelihood function using a canonical parametrization $\varphi=\varphi(\theta)$, which is computed locally at the data point.  This generalizes the method of eliminating nuisance parameters by conditioning or marginalizing to more general contexts.  In Section 2 we give some background notation and introduce the notion of orthogonal parameters.  In Section 3 we illustrate the $p$-value function approach in a simple model with no nuisance parameters.  Profile likelihood and adjustments to profile likelihood are described in Section 4.  Third order $p$-values for problems with nuisance parameters are described in Section 5.  Section 6 describes the classical conditional and marginal likelihood approach.

\section{NOTATION AND ORTHOGONAL PARAMETERS}

We assume our measurement(s) $y$ can be modelled as coming from a probability distribution with density or mass function $f(y;\theta)$, where $\theta=(\psi, \lambda)$ takes values in $R^d$.  We assume $\psi$ is a one-dimensional parameter of interest, and $\lambda$ is a vector of nuisance parameters.  If there is interest in more than one component of $\theta$, the methods described here can be applied to each component of interest in turn.  The likelihood function is
\begin{equation}
L(\theta)=L(\theta;y)=c(y)f(y;\theta);
\end{equation}
it is defined only up to arbitrary multiples which may depend on $y$ but not on $\theta$.  This ensures in particular that the likelihood function is invariant to one-to-one transformations of the measurement(s) $y$.  In the context of independent, identically distributed sampling, where $y=(y_1, \dots , y_n)$ and each $y_i$ follows the model $f(y;\theta)$ the likelihood function is proportional to $\Pi f(y_i;\theta)$ and the log-likelihood function becomes a sum of independent and identically distributed components:
\begin{equation}
\ell(\theta)=\ell(\theta;y)=\Sigma \log f(y_i;\theta)+a(y).
\end{equation}
The maximum likelihood estimate $\hat\theta$ is the value of $\theta$ at which the likelihood takes its maximum, and in regular models is defined by the score equation
\begin{equation}
\ell'(\hat\theta;y)=0.
\end{equation}
The observed Fisher information function $j(\theta)$ is the curvature of the log-likelihood:
\begin{equation}
j(\theta)=-\ell''(\theta)
\end{equation}
and the expected Fisher information is the model quantity
\begin{equation}
i(\theta)=E\{-\ell''(\theta)\}=\int -\ell''(\theta;y)f(y;\theta)dy.
\end{equation}
If $y$ is a sample of size $n$ then $i(\theta)=O(n)$.  

In accord with the partitioning of $\theta$ we partition the observed and expected information matrices and use the notation
\begin{equation}
i(\theta)=\left( \begin{array}{cc} i_{\psi\psi}&i_{\psi\lambda} \\
i_{\lambda\psi}&i_{\lambda\lambda} \end{array}\right )
\end{equation} and
\begin{equation}
i^{-1}(\theta) =\left( \begin{array}{cc} i^{\psi\psi}&i^{\psi\lambda} \\
i^{\lambda\psi}&i^{\lambda\lambda} \end{array}\right ) .
\end{equation}
We say $\psi$ is {\em orthogonal} to $\lambda$ (with respect to expected Fisher information) if $i_{\psi\lambda}(\theta)=0$.  When $\psi$ is scalar a transformation from $(\psi,\lambda)$ to $(\psi,\eta(\psi,\lambda))$ such that $\psi$ is orthogonal to $\eta$ can always be found (Cox and Reid, [1]).  The most directly interpreted consequence of parameter orthogonality is that the maximum likelihood estimates of  orthogonal components are asymptotically independent.

{\bf Example 1:\, ratio of Poisson means}
Suppose $y_1$ and $y_2$ are independent counts modelled as Poisson with mean $\lambda$ and  $\psi\lambda$, respectively.  Then the likelihood function is
$$
L(\psi,\lambda; y_1, y_2)=e^{-\lambda(1+\psi)} \psi^{y_2} \lambda^{y_1+y_2}
$$
and $\psi$ is orthogonal to $\eta(\psi,\lambda)=\lambda(\psi+1)$.  In fact in this example the likelihood function factors as $L_1(\psi)L_2(\eta)$, which is a stronger property than parameter orthogonality.  The first factor is the likelihood for a binomial distribution with index $y_1+y_2$ and
probability of success $\psi/(1+\psi)$, and the second is that for a Poisson distribution with
mean $\eta$.

{\bf Example 2:\, exponential regression}
Suppose $y_i, i=1, \dots, n$ are independent observations, each from an exponential distribution with mean $\lambda\exp(-\psi x_i)$, where $x_i$ is known.
The log-likelihood function is
\begin{equation}
\ell(\psi,\lambda;y)=-n\log\lambda+\psi\Sigma x_i -\lambda^{-1}\Sigma y_i \exp(\psi x_i)
\end{equation}
and $i_{\psi\lambda}(\theta)=0$ if and only if $\Sigma x_i=0$. The stronger property of factorization of the likelihood does not hold.

\section{LIKELIHOOD INFERENCE WITH NO NUISANCE PARAMETERS}

We assume now that $\theta$ is one-dimensional.
A plot of the log-likelihood function as a function of $\theta$ can quickly reveal irregularities in the model, such as a non-unique maximum, or a maximum on the boundary, and can also provide a visual guide to deviance from normality, as the log-likelihood function for a normal distribution is a parabola and hence symmetric about the maximum.  In order to calibrate the log-likelihood function we can use the approximation
\begin{equation}
r(\theta)={\rm sign}(\hat\theta-\theta)[2\{\ell(\hat\theta)-\ell(\theta)\}]^{1/2} \dotsim N(0,1),
\end{equation}
which is equivalent to the result that twice the log likelihood ratio is approximately $\chi_1^2$.  This will typically provide a better approximation than the asymptotically equivalent result that
\begin{equation}
\hat\theta-\theta \dotsim N(0, i^{-1}(\theta))
\end{equation}
as it partially accommodates the potential asymmetry in the log-likelihood function.  These two approximations are sometimes called first order approximations because in the context where the log-likelihood is $O(n)$, we have (under regularity conditions) results such as
\begin{eqnarray}
{\rm Pr}\{r(\theta;y) \le r(\theta;y^0)\} &=& {\rm Pr}\{Z\le r(\theta;y^0)\}
\\
&&\quad\quad\{1+O(n^{-1/2})\} \nonumber 
\end{eqnarray}
where $Z$ follows a standard normal distribution.
It is relatively simple to improve the approximation to third order, i.e. with relative error $O(n^{-3/2})$, using the so-called $r^*$ approximation 
\begin{equation}
r^*(\theta)=r(\theta)+\{1/r(\theta)\}\log\{q(\theta)/r(\theta)\}\sim N(0,1)
\end{equation}
where $q(\theta)$ is a likelihood-based statistic and a generalization of the Wald statistic
$(\hat\theta-\theta)j^{1/2}(\hat\theta)$; see Fraser [2].

\noindent{\bf Example 3: truncated Poisson}

Suppose that $y$ follows a Poisson distribution with mean $\theta=b+\mu$, where $b$ is a background rate that is assumed known.  In this model the $p$-value function can be computed exactly simply by summing the Poisson probabilities.  Because the Poisson distribution is discrete, the $p$-value could reasonably be defined as either 
\begin{equation}
{\rm Pr}(y \le y^0;\theta)
\end{equation} or
\begin{equation}
{\rm Pr}(y  <  y^0;\theta),
\end{equation}
sometimes called the upper and lower $p$-values, respectively.  

For the values $y^0=17$, $b=6.7$, Figure 1 shows the likelihood function as a function of $\mu$ and the $p$-value function $p(\mu)$ computed using both the upper and lower $p$-values.    In Figure 2 we plot the {\em mid} $p$-value, which is 
\begin{equation}
{\rm Pr}(y<y^0)+(1/2){\rm Pr}(y=y^0).
\end{equation}
The approximation based on $r^*$ is nearly identical to the mid-$p$-value; the difference cannot be seen on Figure 2.  Table 1 compares the $p$-values at $\mu=0$.
This example is taken from Fraser, Reid and Wong [3].

\begin{figure}
\centering
\includegraphics[width=3.2in,height=4.2in]{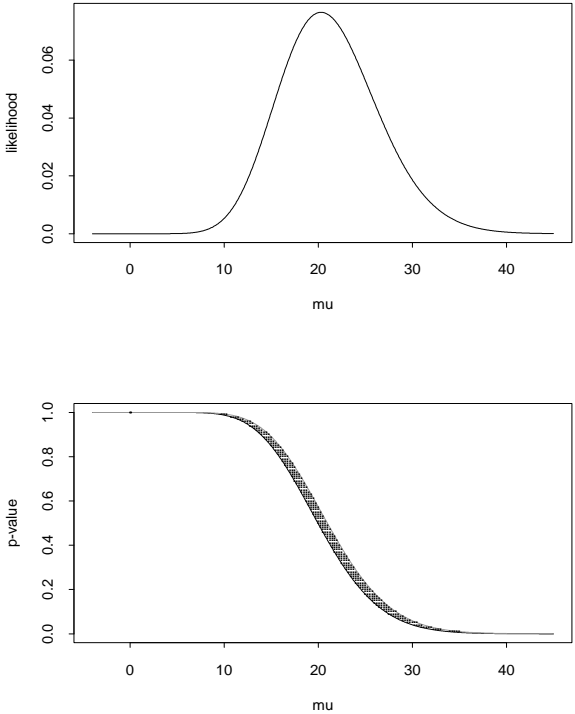}
\caption{The likelihood function (top) and $p$-value function (bottom) for the Poisson model, with $b=6.7$ and $y^0=17$.  For $\mu=0$ the $p$-value interval is $(0.99940, 0.99978)$.}
\end{figure}

\begin{figure}
\centering
\includegraphics[width=3.2in,height=3.2in]{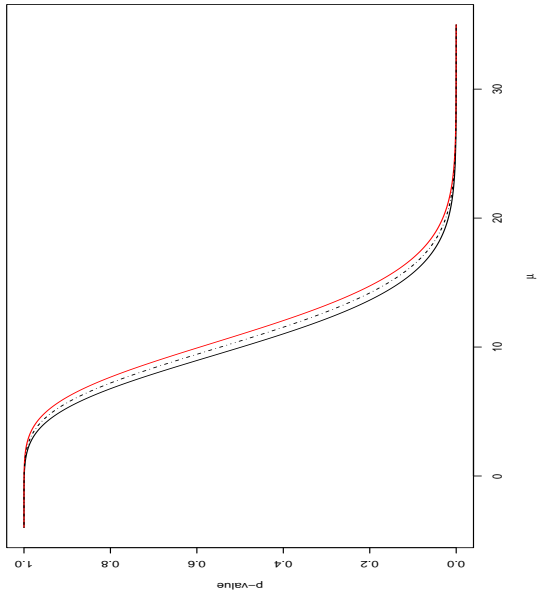}
\caption{The upper and lower $p$-value functions and the mid-$p$-value function for the Poisson model, with $b=6.7$ and $y^0=17$. The approximation based on $\Phi(r^*)$ is identical to the mid-$p$-value function to the drawing accuracy.}
\end{figure}

\begin{table}
\begin{center}
\caption{The $p$-values for testing $\mu=0$, i.e. that the number of observed events is consistent with the background.}
\begin{tabular}{ll} \hline
upper $p$-value &  0.0005993 \\
   lower $p$-value &  0.0002170  \\
   mid $p$-value & 0.0004081\\
$\Phi(r^*)$ & 0.0003779\\
$\Phi(r)$& 0.0004416\\
$\Phi\{(\hat\theta-\theta)\hat j^{1/2}\}$& 0.0062427 \\ \hline
\end{tabular}
\end{center}
\end{table}

\section{PROFILE AND ADJUSTED PROFILE LIKELIHOOD FUNCTIONS}

We now assume $\theta=(\psi,\lambda)$ and denote by $\hat\lambda_\psi$ the restricted maximum likelihood estimate obtained by maximizing the likelihood function over the nuisance parameter $\lambda$ with $\psi$ fixed.  The profile likelihood function is 
\begin{equation}
L_p(\psi)=L(\psi,\hat\lambda_\psi);
\end{equation}
also sometimes called the concentrated likelihood or the peak likelihood.  The approximations of the previous section generalize to 
\begin{equation}
r(\psi)={\rm sign}(\hat\psi-\psi)[2\{\ell_p(\hat\psi)-\ell_p(\psi)\}]^{1/2} 
\dotsim N(0,1),
\end{equation}
and
\begin{equation}
\hat\psi-\psi \dotsim N(0, \{i^{\psi\psi}(\theta)\}^{-1}) .
\end{equation}
These approximations, like the ones in Section 3, are derived from asymptotic results which assume that $n\rightarrow \infty$, that we have a vector $y$ of independent, identically distributed observations, and that  the dimension of the nuisance parameter does not increase with $n$.  Further regularity conditions are required on the model, such as are outlined in textbook treatments of the asymptotic theory of maximum likelihood. In finite samples these approximations can be misleading: profile likelihood is too concentrated, and can be maximized at the `wrong' value.

{\bf Example 4: normal theory regression}
Suppose $y_i=x_i'\beta +\epsilon_i$, where $x_i = (x_{i1}, \dots, x_{ip})$ is a vector of known covariate values, $\beta$ is an unknown parameter of length $p$, and $\epsilon_i$ is assumed to follow a $N(0, \psi)$ distribution.  The maximum likelihood estimate of $\psi$ is
\begin{equation}
\hat\psi=\frac{1}{n} \Sigma (y_i- x_i'\hat\beta)^2
\end{equation}
which tends to be too small, as it does not allow for the fact that $p$ unknown parameters (the components of $\beta$) have been estimated.  In this example there is a simple improvement, based on the result that the likelihood function for $(\beta,\psi)$ factors into
\begin{equation}
L_1(\beta,\psi;\bar y)L_2\{\psi;\Sigma(y_i-x_i'\hat\beta)^2\}
\end{equation}
where $L_2(\psi)$ is  proportional to the marginal distribution of 
$\Sigma(y_i-x_i'\hat\beta)^2$.  Figure 3  shows the profile likelihood and the  marginal likelihood; it is easy to verify that the latter  is maximized at 
\begin{equation}
\hat\psi_m = \frac{1}{n-p} \Sigma (y_i-  x_i'\hat\beta)^2
\end{equation}
which in fact is an unbiased estimate of $\psi$.

\begin{figure}\label{fig3}
\caption{Profile likelihood and marginal likelihood for the variance parameter in a normal theory regression with 21 observations and three covariates (the "Stack Loss" data included in the 
Splus distribution).  The profile likelihood is maximized at a smaller value of $\psi$, and is
narrower; in this case  both the estimate and its estimated standard error are too small.}
\includegraphics[width=3.2in,height=3.2in]{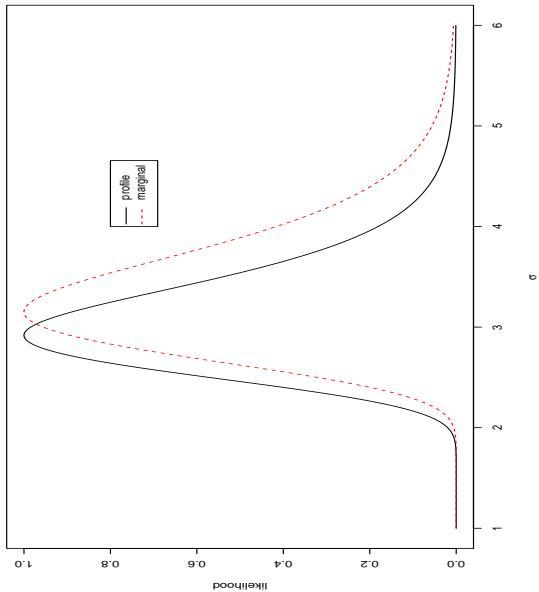}
\end{figure}

{\bf Example 5: product of exponential means}
Suppose we have independent pairs of observations $y_{1i}, y_{2i}$, where
$y_{1i} \sim Exp(\psi\lambda_i) \quad y_{2i} \sim Exp(\psi/\lambda_i) , i=1, \dots, n$.  The limiting normal theory for profile likelihood does not apply in this context, as the dimension of the parameter is not fixed but increasing with the sample size, and it can be shown that 
\begin{equation}
\hat\psi\rightarrow \frac{\pi}{4}\psi
\end{equation}
as $n\rightarrow\infty$ (Cox and Reid [4]).

The theory of higher order approximations can be used to derive a general improvement to the profile likelihood or log-likelihood function, which takes the form
\begin{equation}
\ell_a(\psi)=\ell_p(\psi)+\frac{1}{2}\log |j_{\lambda\lambda}(\psi,\hat\lambda_\psi)| + B(\psi)
\end{equation}
where $j_{\lambda\lambda}$ is defined by the partitioning of the observed information function, and $B(\psi)$ is a further adjustment function that is $O_p(1)$.  Several versions of $B(\psi)$ have been suggested in the statistical literature: we use the one defined in Fraser [5] given by
\begin{equation}
B(\psi)=\displaystyle{-\frac{1}{2} \log |\varphi_\lambda'(\psi,\hat\lambda_\psi)
 j_{\varphi\varphi}(\hat\psi,\hat\lambda) \varphi_\lambda'(\psi,\hat\lambda_\psi)|}.
 \end{equation}
 This depends on a so-called canonical parametrization $\varphi=\varphi(\theta)=\ell_{;V}(\theta;y^0)$ 
which is discussed in Fraser, Reid and Wu [6] and Reid [7].

In the special case that $\psi$ is orthogonal to the nuisance parameter $\lambda$ a simplification of $\ell_a(\psi)$ is available as
\begin{equation}
\ell_{CR}(\psi)=\ell_p(\psi)-\displaystyle{ \frac{1}{2}}\log |j_{\lambda\lambda}(\psi,\hat\lambda_\psi)|
\end{equation}
which was first introduced in Cox and Reid (1987).  The change of sign on $\log|j|$ comes from the orthogonality equations.  In i.i.d. sampling, $\ell_p(\psi)$ is $O_p(n)$, i.e. is the sum of $n$ bounded random variables, whereas $\log|j|$ is $O_p(1)$.  A drawback of $\ell_{CR}$ is that it is not invariant to one-to-one reparametrizations of $\lambda$, all of which are orthogonal to $\psi$.  In contrast $\ell_a(\psi)$ is invariant to transformations $\theta=(\psi,\lambda)$ to $\theta'=(\psi,\eta(\psi,\lambda))$, sometimes called interest-respecting transformations.

{\bf Example 5 continued}
In this example $\psi$ is orthogonal to $\lambda=(\lambda_1, \dots , \lambda_n)$, and 
\begin{equation}
\ell_{CR}(\psi)= -(3n/2)\log\psi -(2/\psi)\Sigma \surd(y_{1i}y_{2i}) .
\end{equation}
The value that maximizes $\ell_{CR}$ is 'more nearly consistent' than the maximum likelihood estimate as $\hat\psi_{CR}\longrightarrow (\pi/3)\psi$.

\section{$P$-VALUES FROM PROFILE LIKELIHOOD}

The limiting theory for profile likelihood gives first order approximations to $p$-values, such as
\begin{equation}
p(\psi)\doteq\Phi(r_p)
\end{equation}
and
\begin{equation}
p(\psi)\doteq \Phi\{(\hat\psi-\psi)j_p^{1/2}(\hat\psi)\}
\end{equation}
although the discussion in the previous section suggests these may not provide very accurate approximations.  As in the scalar parameter case, though, a much better approximation is available using $\Phi(r^*)$ where
\begin{equation}\label{rstar}
r^*(\psi)=r_p(\psi)+1/\{r_p(\psi)\}\log\{Q(\psi)/r_p(\psi)\}
\end{equation} 
where $Q$ can also be derived from the likelihood function and a function $\varphi(\theta, y^0)$ as
$$Q= (\hat\nu-\hat\nu_\psi) \hat\sigma_\nu^{-1/2}$$
where
\begin{eqnarray*}
\nu(\theta ) &=& e^T_\psi \varphi (\theta ) \ , \nonumber \\
e_\psi &=& \psi_{\varphi'} (\hat\theta_\psi) / 
| \psi_{\varphi'} (\hat\theta_\psi ) | \ , \nonumber \\
\hat\sigma^2_\nu &=& |j_{(\lambda\lambda)} (\hat\theta_\psi) | / 
|j_{(\theta\theta)} (\hat\theta)| \ ,  \\
|j_{(\theta\theta)} (\hat\theta)| &=& |j_{\theta\theta} (\hat\theta) |
| \varphi_{\theta'} (\hat\theta) |^{-2} \ , \\
|j_{(\lambda\lambda)} (\hat\theta_\psi )| &=& |j_{\lambda\lambda} (\hat\theta_\psi) |
| \varphi_{\lambda'} (\hat\theta_\psi) |^{-2} \ . 
\end{eqnarray*}
The derivation is described in Fraser, Reid and Wu [6] and  Reid [7].  The key ingredients are the log-likelihood function $\ell(\theta)$ and a reparametrization $\varphi(\theta)=\varphi(\theta;y^0)$, which is defined by using an approximating model at the observed data point $y^0$; this approximation in turn is based on a conditioning argument.  A closely related approach is due to Barndorff-Nielsen; see Barndorff-Nielsen and Cox [8, Ch.~7], and the two approaches are compared in [7].

{\bf Example 6: comparing two binomials}
Table 2 shows the employment history of men and women at the Space Telescope Science Institute, as reported in {\it Science} Feb 14 2003.  We denote by $y_1$ the number of males who left and model this as a Binomial with sample size 19 and probability $p_1$; similarly the number of females who left, $y_2$, is modelled as Binomial with sample size 7 and probability $p_2$.  We write the parameter of interest
\begin{equation}
\psi=\log\displaystyle{\frac{p_1(1-p_2)}{p_2(1-p_1)}}.
\end{equation}
The hypothesis of interest is $p_1=p_2$, or $\psi=0$.  The $p$-value function for $\psi$ is plotted in Figure 4.  The $p$-value at $\psi=0$ is 0.00028 using the normal approximation to $r_p$, and is 0.00048 using the normal approximation to $r^*$.  Using Fisher's exact test gives a mid $p$-value of 0.00090, so the approximations are anticonservative in this case.

\begin{table}
\caption{Employment of men and women at the  Space Telescope
Science Institute, 1998--2002 (from \emph{Science}\ magazine, Volume 299, page 993, 14 February
2003).}
\begin{center}
\begin{tabular}{lccc}
\\
\hline
&Left&Stayed&Total\\
\cline{1-4}
Men&1&18&19\\
Women&5&2&7\\
\cline{1-4}
Total&6&20&26\\
\hline
\end{tabular}
\end{center}
\end{table}

\begin{figure}
\centering
\includegraphics[width=3.2in,height=3.2in]{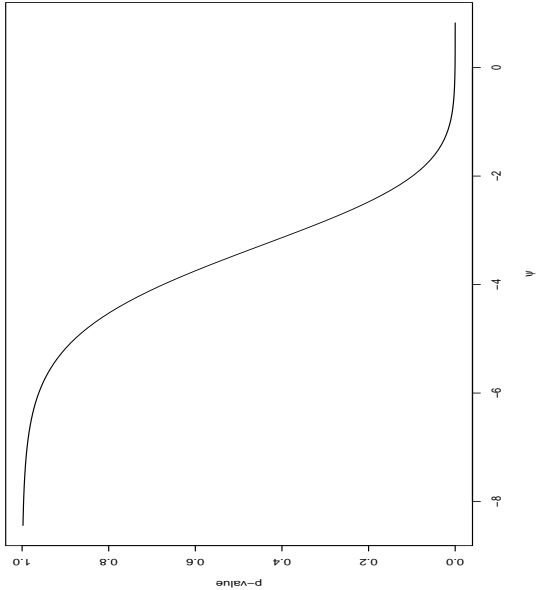}
\caption{The $p$-value function for the log-odds ratio, $\psi$, for the data of Table II.  The value $\psi=0$ corresponds to the hypothesis that the probabilities of leaving are equal for men and women.}
\end{figure}

{\bf Example 7: Poisson with estimated background}
Suppose in the context of Example 3 that we allow for imprecision in the background, replacing $b$ by an unknown parameter $\beta$ with estimated value $\hat\beta$.  We assume that the background estimate is obtained from a Poisson count $x$, which has mean $k\beta$, and the signal measurement is an independent Poisson count, $y$, with mean $\beta+\mu$.  We have $\hat\beta=x/k$ and ${\rm var}{\hat\beta}=\beta/k$, so the estimated precision of the background gives us a value for $k$.  For example, if the background is estimated to be $6.7\pm 2.1$ this implies a value for $k$ of $6.7/(2.1)^2\doteq 1.5$.  Uncertainty in the standard error of the background is ignored here.  We now outline the steps in the computation of the $r^*$ approximation (\ref{rstar}).

The log-likelihood function based on the two independent observations $x$ and $y$  is 
\begin{equation}
\ell(\beta,\mu)=x\log(k\beta)-k\beta +y\log(\beta+\mu)-\beta-\mu
\end{equation}
with canonical parameter $\varphi=(\log\beta,\log(\beta+\mu))'$.

Then
\begin{equation}
\varphi_{\theta'}(\theta)=\frac{\partial\varphi(\theta)}{\partial\theta'}= \left (
\begin{array}{cc}
0&1/\beta \\ 1/(\beta+\mu) & 1/(\beta+\mu) \end{array}\right ),
\end{equation}
\begin{equation}
\varphi^{-1}_{\theta'}=\left ( \begin{array}{cc}
-\beta & \beta+\mu \\ -\beta &0\end{array} \right )
\end{equation}
from which
\begin{equation}
\psi_{\varphi'}=(-\beta, \beta+\mu).
\end{equation}
Then we have
\begin{eqnarray}
\chi(\hat\theta)&=&\frac{-\hat\beta_\mu \log(\hat\beta) + (\hat\beta_\mu+\mu) \log(\hat\beta+\hat\mu)} {\surd\{\hat\beta_\mu^2+(\hat\beta_\mu+\mu)^2\}} \\
\chi(\hat\theta_\psi)&=&\frac{-\hat\beta_\mu \log(\hat\beta_\mu) + (\hat\beta_\mu+\mu) \log(\hat\beta_\mu+\mu)}
{\surd\{\hat\beta_\mu^2+(\hat\beta_\mu+\mu)^2\}} ,
\end{eqnarray}
\begin{eqnarray}
|j_{(\theta\theta)}(\hat\theta)| &=& y_1y_2 =k/\hat\beta(\hat\beta+\hat\mu) \\
|j_{(\lambda\lambda)}(\hat\theta_\psi)|&=&\frac{y_1(\hat\beta_\mu+\mu)^2 + y_2\hat\beta_\mu^2}{(\hat\beta_\mu+\mu)^2 +\hat\beta_\mu^2}
\end{eqnarray}
and finally
\begin{eqnarray}
Q&=&\left\{(\hat\beta_\mu+\mu)\log\left(\frac{\hat\beta+\hat\mu}{\hat\beta_\mu+\mu}\right ) - \hat\beta_\mu\log\frac{\hat\beta}{\hat\beta_\mu}\right\} \nonumber \\
&&\quad \frac{\{k\hat\beta(\hat\beta+\hat\mu)\}^{1/2}}
{\{k\hat\beta(\hat\beta_\mu+\mu)^2+(\hat\beta+\hat\mu)\hat\beta_\mu^2\}^{1/2}}.
\end{eqnarray}
The likelihood root is
\begin{eqnarray}
r&=&{\rm sign}(Q)\surd[2\{\ell(\hat\beta,\hat\mu)-\ell(\hat\beta_\mu,\mu)\}]\\
&=& {\rm sign}(Q)\surd(2[k\hat\beta\log\{\hat\beta/\hat\beta_\mu\}) + (\hat\beta+\hat\mu) 
\nonumber \\
&& \quad \log\{(\hat\beta+\hat\mu)/(\hat\beta_\mu+\mu)\} \nonumber  \\
&& -k(\hat\beta-\hat\beta_\mu)-\{\hat\beta+\hat\mu-(\hat\beta_\mu+\mu)\}]).
\end{eqnarray}
The third order approximation to the $p$-value function is $1-\Phi(r^*)$, where
\begin{equation}
r^*=r+(1/r)\log(Q/r).
\end{equation}

Figure 5 shows the $p$-value function for $\mu$ using the mid-$p$-value function from the Poisson with no adjustment for the error in the background, and the $p$-value function from $1-\Phi(r^*)$.  The $p$-value for testing $\mu=0$ is 0.00464, allowing for the uncertainty in the background, whereas it is 0.000408 ignoring this uncertainty.  

The hypothesis $Ey=\beta$ could also be tested by modelling the mean of $y$ as $\nu\beta$, say, and testing the value $\nu=1$.  In this formulation we can eliminate the nuisance parameter exactly by using the binomial distribution of $y$ conditioned on the total $x+y$, as described in example 1.    
This gives a mid-$p$-value of 0.00521.  The computation is much easier than that outlined above, and seems quite appropriate for testing the equality of the two means.  However if inference about the mean of the signal is needed, in the form of a point estimate or confidence bounds, then the formulation as a ratio seems less natural at least in the context of HEP experiments.  A more complete comparison of methods for this problem is given in Linnemann [8].

\begin{figure*}
\centering
\includegraphics[width=135mm]{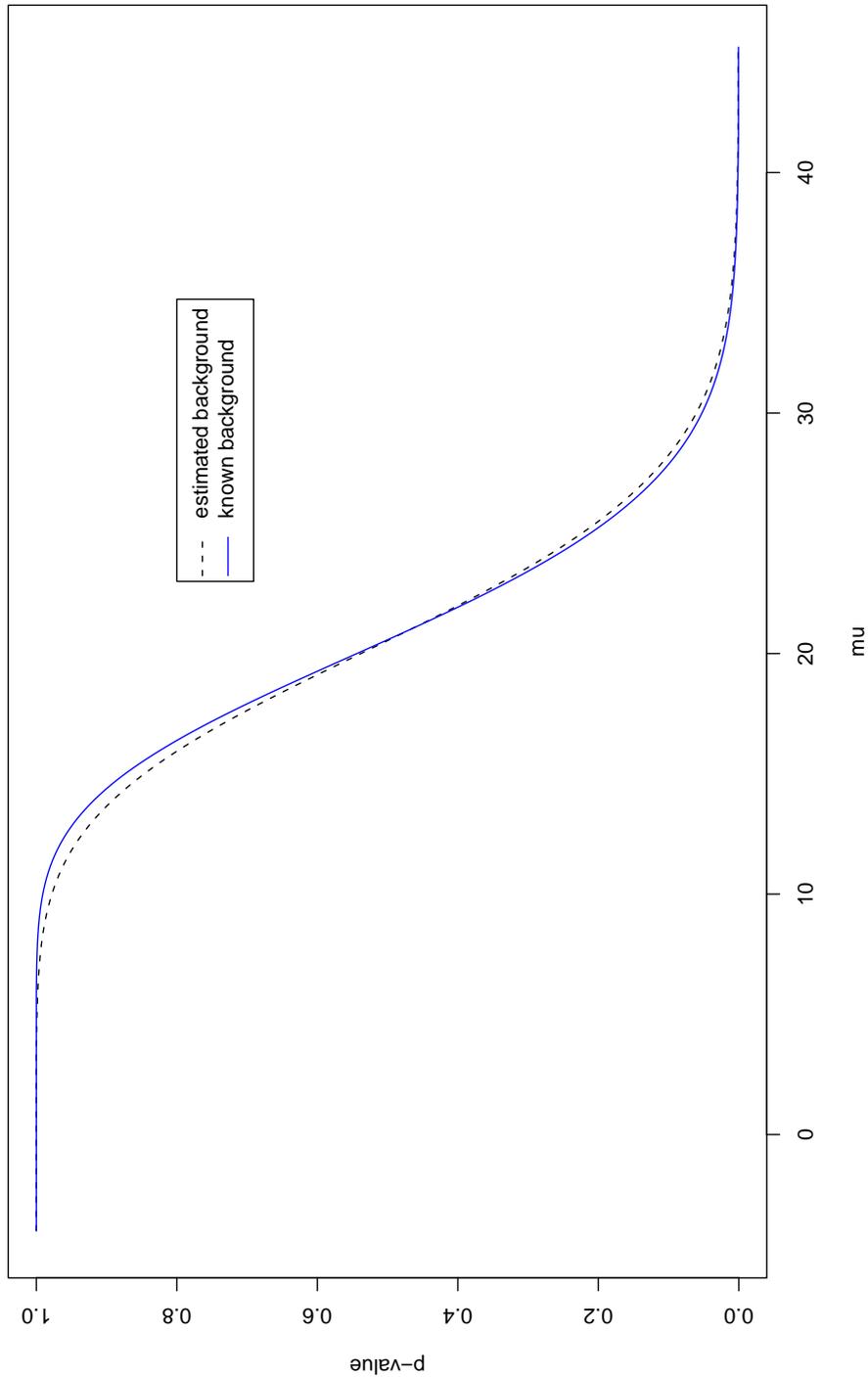}
\caption{Comparison of the $p$-value functions computed assuming the background is known and using the mid-$p$-value with the third order approximation allowing a background error of $\pm 1.75$.}
\end{figure*}

\section{CONDITIONAL AND MARGINAL LIKELIHOOD}

In special model classes, it is possible to 
eliminate nuisance parameters by either
 {\em conditioning} or {\em marginalizing}.  The conditional or marginal likelihood then gives essentially
  exact inference for the parameter of interest, if this likelihood can itself be computed exactly.   In Example 1 above, $L_1$ is the density for $y_2$ conditional on $y_1+y_2$, so is a conditional likelihood for $\psi$.  This is an example of the more general class of linear exponential families:
\begin{equation}\label{expon}
f(\underline y;\psi,\lambda)=\exp\{\psi s(\underline y) + \lambda't(\underline y)-c(\psi,\lambda)-d(\underline y)\};
\end{equation}
in which
\begin{equation}\label{fcond}
f_{cond}(s\mid t;\psi) = \exp\{\psi s -C_t(\psi)-D_t(s)\}
\end{equation}
defines the conditional likelihood.  The comparison of two binomials in Example 6 is in this class, with $\psi$ as defined at (30) and $\lambda=\log\{p_2/(1-p_2)\}$.  The difference of two Poisson means, in Example 7, cannot be formulated this way, however, even though the Poisson distribution is  an exponential family, because the parameter of interest $\psi$ is not a component of the canonical parameter.

It can be shown  that in models of the form (\ref{expon}) the  log-likelihood 
$\ell_a(\psi)=\ell_p(\psi)+(1/2)\log|j_{\lambda\lambda}|$ approximates the conditional log-likelihood $\ell_{cond}(\psi)=\log f_{cond}(s\mid t;\psi)$, and that  
\begin{equation}
p(\psi)=\Phi(r^*)
\end{equation}
where
\begin{eqnarray*}
r^*&=&r_a+\displaystyle{\frac{1}{r_a}}\log(\displaystyle{\frac{Q}{r_a}})\\
r_a&=& \pm [2\{\ell_a(\hat\psi_a)-\ell_a(\psi)\}]^{1/2}\\
Q &=& (\hat\psi_a-\psi)\{j_a(\hat\psi)\}^{1/2}
\end{eqnarray*}
approximates the $p$-value function with relative error $O(n^{-3/2})$ in i.i.d. sampling.  An asymptotically equivalent approximation based on the profile log-likelihood is 
\begin{equation}
p(\psi)=\Phi(r^*)
\end{equation}
where
\begin{eqnarray*}
r^*&=&r_p+\displaystyle{\frac{1}{r_p}}\log(\displaystyle{\frac{Q}{r_p}})\\
r_p&=& \pm [2\{\ell_p(\hat\psi)-\ell_p(\psi)\}]^{1/2}\\
Q &=& (\hat\psi-\psi)\{j_p(\hat\psi)\}^{1/2}
\frac{|j_{\lambda\lambda}(\psi,\hat\lambda_\psi)|^{1/2}}
{|j_{\lambda\lambda}(\hat\psi,\hat\lambda)|^{1/2}} .
\end{eqnarray*}
In the latter approximation an adjustment for nuisance parameters is made to $Q$, whereas in the former the adjustment is built into the likelihood function.  Approximation (46) was used in Figure 3.

A similar discussion applies to the class of transformation models, using marginal approximations.  Both classes are reviewed in
Reid [9]. 

\begin{acknowledgments}
The authors wish to thank Anthony Davison and Augustine Wong for helpful discussion.  This research was partially supported by the Natural Sciences and Engineering Research Council.
\end{acknowledgments}

\bigskip


\end{document}